\documentclass[final,5p,times]{elsarticle}
\usepackage{amsmath,amsfonts,graphics,graphicx,amssymb}

\usepackage{hyperref}
\usepackage{a4wide}
\usepackage{graphicx}
\usepackage{subfigure}
\usepackage{epsfig}
\usepackage{atlasphysics}
\usepackage{mathptmx}\usepackage{amsmath}
\usepackage{amssymb}
\usepackage{bm}

\newcommand{\uones}{\ensuremath{\Upsilon(1S)}}
\newcommand{\utwos}{\Upsilon(2S)}
\newcommand{\uthrees}{\Upsilon(3S)}
\newcommand{\ptmm}{p_T^{\mu\mu}}
\newcommand{\ymm}{y^{\mu\mu}}
\newcommand{\mmm}{m^{\mu\mu}}

\journal{Physics Letters B}

\begin{document}

\begin{frontmatter}
 
\title{Measurement of the {$\ensuremath{\uones}$} Production Cross-Section in $pp$ Collisions at {$\sqrt{s}=7$}~TeV in ATLAS}
 
\author{The ATLAS Collaboration}
 
\begin{abstract}
A measurement of the cross-section for $\uones \to \mu^+\mu^-$ production in proton-proton collisions at centre of mass energy of 7 TeV is presented. The cross-section is measured as a function of the $\uones$ transverse momentum in two bins of rapidity, $|y^{\uones}|<1.2$ and $1.2<|y^{\uones}|<2.4$. 
The measurement requires that both muons have transverse momentum $p_T^{\mu}>4$~GeV and pseudorapidity $|\eta^{\mu}|<2.5$ in order to reduce theoretical uncertainties on the acceptance, which depend on the poorly known polarization.
The results are based on an integrated luminosity of 1.13~pb$^{-1}$, collected with the ATLAS detector at the Large Hadron Collider. The cross-section measurement is compared to theoretical predictions: it agrees to within a factor of two with a prediction based on the NRQCD model including colour-singlet and colour-octet matrix elements as implemented in {\sc Pythia} while it disagrees by up to a factor of ten with the next-to-leading order prediction based on the colour-singlet-model.
\end{abstract}
 
\end{frontmatter}

\section{Introduction}
The production of $J/\psi$ and $\Upsilon$ mesons has been studied since their discovery in the 1970s~\cite{jpsi,ups}, 
and even today there is no conclusive coherent theoretical picture of
J/$\psi$ and $\Upsilon$ hadroproduction. There are major questions
associated with the consistency of measurements made at the Tevatron,
HERA and fixed target experiments~\cite{review}. In particular it is difficult to reconcile the cross-section measurements with those of the spin alignment~\cite{review}, and there are significant disagreements
between the two Tevatron experiments on the measurement of the spin alignment in the case of $\Upsilon$
production~\cite{cdf,d0}. It is thus important for the LHC
experiments to measure the production of these mesons in order to shed further light on the puzzle.

The CMS collaboration has recently presented a measurement of
$\Upsilon$ production~\cite{cms}, correcting for the detector acceptance of the two daughter muons in both angular and momentum range.
This approach was also adopted in the ATLAS
$J/\psi$ cross-section publication~\cite{jpsipaper}. Correction of the data
in this way introduces uncertainties due to the poorly known spin alignment of
the $\Upsilon$ (or $J/\psi$). Note, that CMS
quote the measurement for a wide variety of assumptions on the spin alignment, and the ATLAS J$/\psi$ measurement takes the 
differences between the corrections obtained with different spin alignments as an additional systematic uncertainty.

\begin{sloppypar}
With that in mind, a somewhat different approach to the
measurement of the $\uones$ production cross-section is adopted in this analysis.
The result is presented as a function of transverse momentum, $p_T$, and
rapidity, $y$, of the $\uones$, corrected for detector response and
efficiencies but defined within a restricted range of muon kinematics where both muons have $p_T^\mu$ greater than 4 GeV and absolute pseudorapidity\footnote{ATLAS uses a right-handed coordinate system with its origin at the nominal interaction point (IP) in the centre of the detector and the $z$-axis along the beam pipe. The $x$-axis points from the IP to the centre of the LHC ring, and the $y$ axis points upward. Cylindrical coordinates $(r,\phi)$ are used in the transverse plane, $\phi$ being the azimuthal angle around the beam pipe. The pseudorapidity is defined in terms of the polar angle $\theta$ as $\eta=-\ln\tan(\theta/2)$.}, $|\eta^\mu|$, less than 2.5. The relative fraction of  $\uones$ mesons where both muons fulfill the kinematic requirements compared to all $\uones$ mesons depends strongly on the spin alignment of the $\uones$. For instance, assuming full transverse or longitudinal spin alignment versus unpolarised production changes this fraction by typically 30\%.  By quoting the measurement in a restricted region of phase space
where muons are detected, uncertainties due to the $\Upsilon$ spin alignment
on the measurement are almost eliminated so that the quoted cross-section is free of any assumptions about this property.
\end{sloppypar}

A single muon trigger with a threshold of $p_T^\mu>4$~GeV is used. This limits the dataset used for this measurement to the low luminosity periods of 2010 corresponding to an integrated luminosity of $1.13 \pm 0.04$ ~pb$^{-1}$.

In the following sections, a brief description of the ATLAS detector is given with emphasis on the aspects most relevant to this analysis. Next the measurement strategy is outlined, followed by a description of the Monte Carlo simulation used.
The event selection and the determination of the number of $\uones$ events are then described before the results are presented, and a conclusion is given.

\section{The ATLAS Detector}
\label{sec:det}

The ATLAS detector~\cite{atlas} consists of an inner tracker, a calorimeter and a muon system. The inner detector (ID) directly surrounds the interaction point; it includes a silicon pixel detector (Pixel), a silicon strip detector (SCT) and a transition radiation tracker (TRT), and is embedded in an solenoidal 2~T magnetic field. 
The ID covers the range $|\eta|<2.5$ and is enclosed by a calorimeter system containing electromagnetic and
hadronic sections. The calorimeter is surrounded by a large muon spectrometer (MS) inside an air-core toroid magnet system
which contains a combination of monitored drift tubes
(MDT) and cathode strip chambers (CSC), designed to provide precise
position measurements in the bending plane and covering the range $|\eta|<2.0$ and $2.0<|\eta|<2.7$, respectively. 
In addition, resistive plate chambers (RPC) and thin gap chambers (TGC) with a coarse position resolution but 
a fast response time are used primarily to trigger muons in the rapidity ranges $|\eta|<1.05$ and $1.05 <|\eta|< 2.4$, respectively.
Momentum measurements in the MS are based on track segments formed separately in at least two of the three station layers of the MDT and the CSC. The RPC and TGC are used to improve the pattern recognition and track reconstruction in the non-bending plane. They do not improve the position measurement in the bending plane.

The first level muon trigger looks for hit coincidences within different RPC or TGC detector layers inside programmed geometrical 
windows that define the muon $p_T$ and provide a rough estimate of their positions~\cite{triggerpaper}. The lowest available $p_T$ threshold is used for this analysis. In addition, muons are required to pass a high-level trigger selection similar to that of the offline reconstruction and a transverse momentum threshold of $p_T^\mu>4$~GeV.

\section{Outline of the Measurement}
\label{sec:outline}

The differential $\uones$ cross-section is given as 
\begin{eqnarray}
\frac{{\rm d}^{2}\sigma}{{\rm d}p_{T}{\rm d}y} \times {\rm BR}(\Upsilon(1S)\to \mu^+\mu^-) & = & \nonumber\\
\frac{N_{\uones}}{{\int{\cal L}{\rm d}t} \times \Delta p_{T} \times \Delta y},
\label{eq:xsec}
\end{eqnarray}
where $\int{\cal L}{\rm d}t$ is the integrated luminosity, and $\Delta p_{T}$ and $\Delta y$ are the
bin sizes in $p_T^{\uones}$ and $y^{\uones}$, respectively. 
$N_{\uones}$ is the corrected number of $\uones$ mesons. It is determined with an unbinned maximum likelihood fit to the dimuon
mass distribution after applying a weight to each candidate that is the inverse of its selection efficiency as described in Section~\ref{sec:evsel}.
The cross-section is defined within the fiducial cuts $p_T^\mu>4$~GeV and $|\eta^\mu|<2.5$ on both muons, where the $\mu$ kinematics are those before any final state QED radiation.

The key aspects of this measurement are the efficiency
determination and the fit to extract $N_{\uones}$. These are
described in detail in Sections~\ref{sec:evsel} and~\ref{sec:upsfit}, respectively.

The effect of bin migrations due to finite detector resolution and final state photon radiation has
been studied. Given the good muon momentum resolution  
of $\sigma(p_T)/p_T<0.5\%$ in the momentum range relevant for this analysis~\cite{resmuonconfnote} and the
relatively coarse binning used for this measurement, the bin
migrations due to detector resolution and final state radiation are smaller than 2\% in all bins. This 2\% also accounts for the migrations across the $p_T^\mu$ and $\eta^\mu$ cuts. This small effect is not corrected for and therefore considered as part of the systematic uncertainty, which is discussed in Section~\ref{sec:sys}. Any residual impacts of the assumption on the $\Upsilon$ polarization have also been assessed and are within this 2\% uncertainty.

\section{Monte Carlo Simulation}
\label{sec:mc}

In this analysis, all efficiency factors are determined directly from the data. The differential cross-section expressed in Eq.~\ref{eq:xsec} does not need any large acceptance corrections which would require
a detailed modelling of the kinematic properties of the events. Monte Carlo (MC) simulation is only used to  
construct templates for the likelihood fits to the dimuon mass distribution, and to assess the corrections 
due to migrations between $p_T^{\uones}$ and $y^{\uones}$ bins.

\begin{sloppypar}
MC events are generated using \textsc{Pythia6}\,\cite{pythia}
with the ATLAS MC09 tune\,\cite{MC09} and MRST LO$^\star$\,\cite{MRSTLO} parton distribution functions.
They are simulated with the ATLAS simulation framework~\cite{atlassim} using \textsc{Geant4}\ \cite{g4} 
and fully reconstructed with the same software that is used to process the data from the detector. 

For the $\Upsilon$ MC samples, \textsc{Pythia6}'s implementation of $\Upsilon$ production subprocesses 
using the non-relativistic QCD (NRQCD)~\cite{nrqcd} framework and the 
parameters recommended in Ref.~\cite{nason} are used. In this model, quarkonium is produced in
both a colour-singlet and a colour-octet state, and evolves non-perturbatively into physical quarkonium.
Each $\Upsilon(nS)$ state is generated separately and includes direct
production from the hard interaction, as well as production 
through radiative feed-down from $\chi_{b}(nP)\to\Upsilon(nS)\gamma$ decays.
The samples are generated without polar or azimuthal anisotropy in the decay of the $\Upsilon$ 
(the default in \textsc{Pythia}).

Background contributions come mainly from open production of charm and bottom quarks with subsequent decay 
of the $c$- or $b$-hadron to a muon. 
A further, much smaller contribution comes from Drell-Yan production.
These continuum backgrounds are described using the minimum-bias processes in \textsc{Pythia}.
In order to avoid double-counting, any generated minimum bias events containing $\Upsilon$-mesons are explicitly removed from the sample. This background sample is only used to estimate a systematic 
uncertainty on the background modelling.
\end{sloppypar}

In all samples final state QED radiation is considered using {\sc Photos}~\cite{photos} interfaced to {\sc Pythia}.

\section{Event Selection and Efficiency Determination}
\label{sec:evsel}

Selected events are first requested to satisfy a single muon trigger with a threshold of $p_T^\mu>4$~GeV. 
Then two offline muons are required with $p_T^\mu>4$~GeV and $|\eta^\mu|<2.5$. 

Reconstructed muons that combine a track reconstructed in the MS with a track reconstructed 
in the ID are referred to as {\it combined muons}~\cite{atlas}. 
In order to recover efficiency for muons with low momenta, {\it tagged muons} 
extrapolate an ID track to the muon system and attach MS track segments
that are not associated to any MS track. For both categories of muons in this analysis, the 
kinematic properties of the muons are solely determined from the
parameters of the ID tracks associated with the muons.
At least one of the two offline muons must be a combined muon, 
and at least one of them must match geometrically to a trigger muon. At low dimuon $p_T$ 
about 60\% of the events have two combined muons, and at high $p_T$ this fraction increases to about $90\%$.
Since the measurement is restricted to $|y^{\uones}|<2.4$ and $p_T^\mu>4$~GeV at least one of
the two decay muons is within the trigger pseudorapidity acceptance.

Both muon tracks are required to have at least one pixel hit and six SCT hits. 
Since the $\Upsilon$ is produced promptly, background from heavy flavour decays is suppressed by requiring the muons to originate from
the primary vertex. Cuts requiring $|d_0|<150$~$\mu$m and $|z_0|\sin\theta<1.5$~mm are applied where $d_0$ ($z_0$) is the impact
 parameter with respect to the event vertex in the transverse  (longitudinal) direction. The event vertex is required to have at
 least three associated tracks to reject background due to cosmic ray muons. The two selected muons are further required to have opposite electric charge.

The muon trigger and reconstruction efficiencies are determined
using the high statistics $J/\psi$ data sample~\cite{muoneff} as a function
of $p_T^\mu$ and $\eta^\mu$. For a single muon, the trigger efficiency varies between 40\% and 90\% 
over the relevant $p_T^\mu$ and $\eta^\mu$ range~\cite{muoneff,triggerpaper}. 
Using the single muon trigger, this results in an average trigger efficiency for the selected dimuon events between 80\% and
95\% in any measurement bin. 
Within the kinematic range considered in this measurement, the offline muon reconstruction 
efficiency varies with $p_T^\mu$ and $\eta^\mu$ between 80\% and 100\%. The small gaps in the muon acceptance at certain $\eta$ regions (mostly at $|\eta| \approx 0$ and $|\eta| \approx 1.3$) are corrected for as part of the efficiency correction assuming the nearly flat $\eta$ dependence predicted by the simulation, and represent a small fraction of the total angular range. The efficiency of the pixel and SCT hit requirements has been measured using $J/\psi$ mesons as $99.5\pm 0.5\%$ per track, and the efficiency of the primary vertex requirement is $>99.9\%$. 
The $z_0\sin\theta$ cut efficiency is nearly 100\% in both data and MC. 

The efficiencies of the transverse impact parameter cut are determined using two independent methods. The impact parameter
resolution is sensitive to the alignment of the ID and to the multiple scattering in the ID volume. 
The former dominates the resolution at high $p_T^\mu$ while the latter dominates at low $p_T^\mu$. 
The primary method to determine the resolution uses muons from $J/\psi$ decays and fits the
impact parameter distribution using templates constructed from prompt and non-prompt $J/\psi$ MC.
In order to allow for small deviations of the observed resolution with respect to the MC,
an additional $d_0$ resolution smearing parameter is introduced.
This smearing parameter is of order 10$\mu$m due to imperfections in 
the material description and tracker alignment.
The efficiency determined with this method agrees well with the simulation and
is about 99.5\% in the central region, decreasing to 96.5\% at the highest $\eta^\mu$ and lowest $p_T^\mu$.
The second method uses $Z$-bosons to determine the impact parameter resolution
at high $p_T^\mu$, and translates it to the low $p_T^\mu$ region using the known dependence
of the resolution on $p_T^\mu$ also taking into account the uncertainty on the material distribution within the ID~\cite{mbpaper}.
The efficiencies measured using the $J/\psi \to \mu^+\mu^-$ decays are used for the central value, and the difference between the two methods is considered as a systematic uncertainty on the muon impact parameter cut efficiency.
It is less than $1\%$ for $|\eta^\mu|<2$, and $1.5-2.5\%$ for $|\eta^\mu|>2$ depending on $p_T^\mu$. 

The efficiencies are accounted for in the cross-section measurement by
applying a weight to each candidate. 
The weight is given by the inverse of the event selection efficiency, $w=1/\epsilon_{\mu\mu}$, which is evaluated from the single-muon trigger and reconstruction efficiencies as

\begin{equation}
\label{eq:efftotal}
\epsilon_{\mu\mu} = \epsilon_{trig}\epsilon_{reco}
\end{equation}
where the individual terms are given by 
\begin{equation}
\epsilon_{trig} = 1 - (1-\epsilon^+_{trig})(1-\epsilon^-_{trig})
\end{equation}
\begin{equation}
\epsilon_{rec} = \epsilon^+_{cb||tag} \epsilon^-_{cb||tag}- \epsilon^-_{tag} \epsilon^+_{tag}
\label{eq:effrec}
\end{equation}

Here $\epsilon^+_{trig}$ and $\epsilon^-_{trig}$ are the trigger efficiencies for a positively and negatively charged muon, respectively.
$\epsilon^+_{cb||tag}$ ($\epsilon^-_{cb||tag} )$ and $\epsilon^+_{tag}$ ($\epsilon^-_{tag} )$ give the single muon reconstruction efficiencies for a combined or tagged or just a tagged muon, respectively, for positively (negatively) charged muons.
For each muon the efficiency is determined as a function of $p_T$, $\eta$ and the electric charge.
The final weights span the range from about 1.4 at low $p_T$ to 1.1 at high $p_T$.

\section{Determination of the Number of {\boldmath $\uones$} Events}
\label{sec:upsfit}

\begin{figure*}[htbp]
\begin{center}
\subfigure[]{\includegraphics[width=0.45\textwidth]{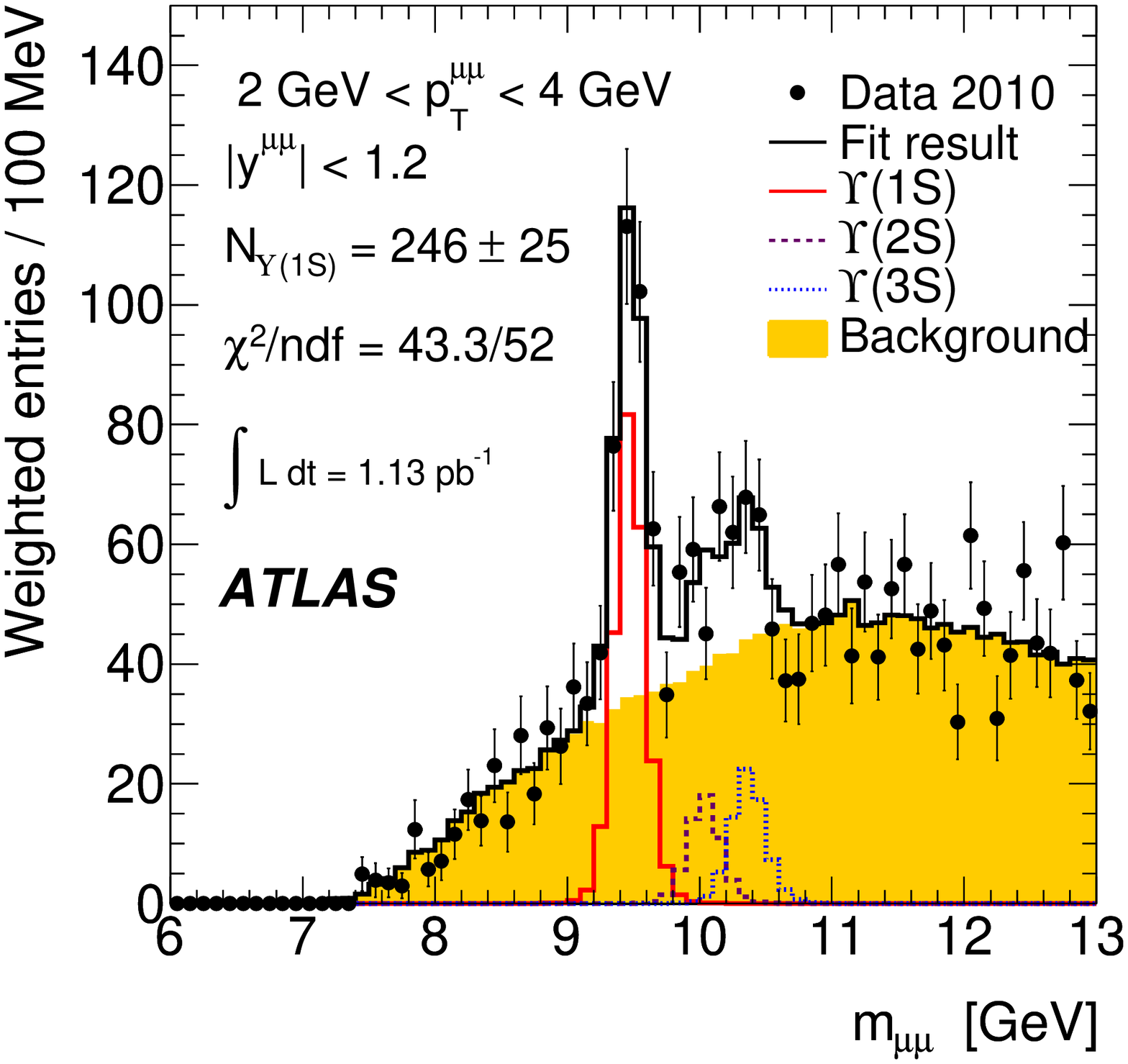}}
\subfigure[]{\includegraphics[width=0.45\textwidth]{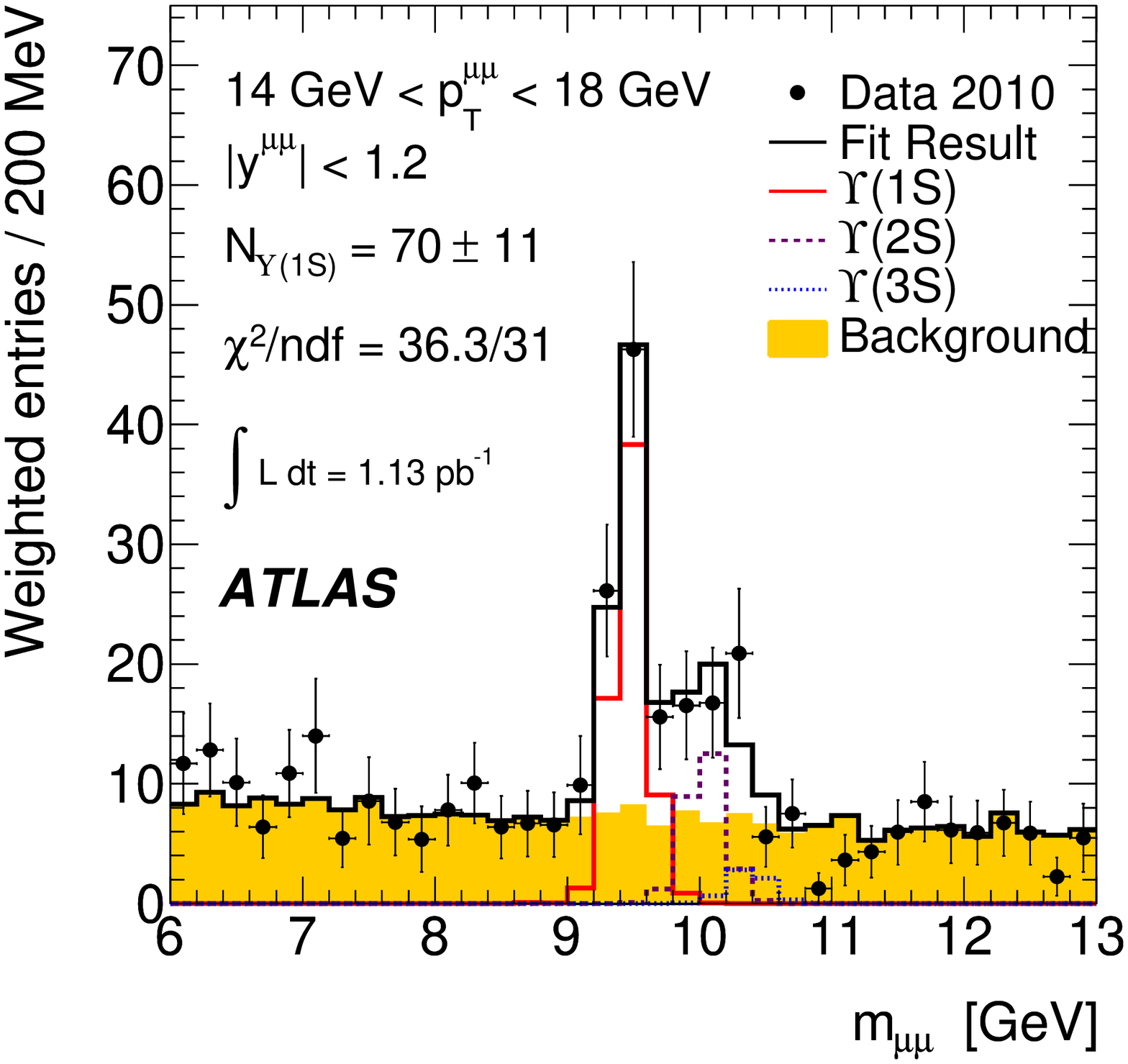}}
\subfigure[]{\includegraphics[width=0.45\textwidth]{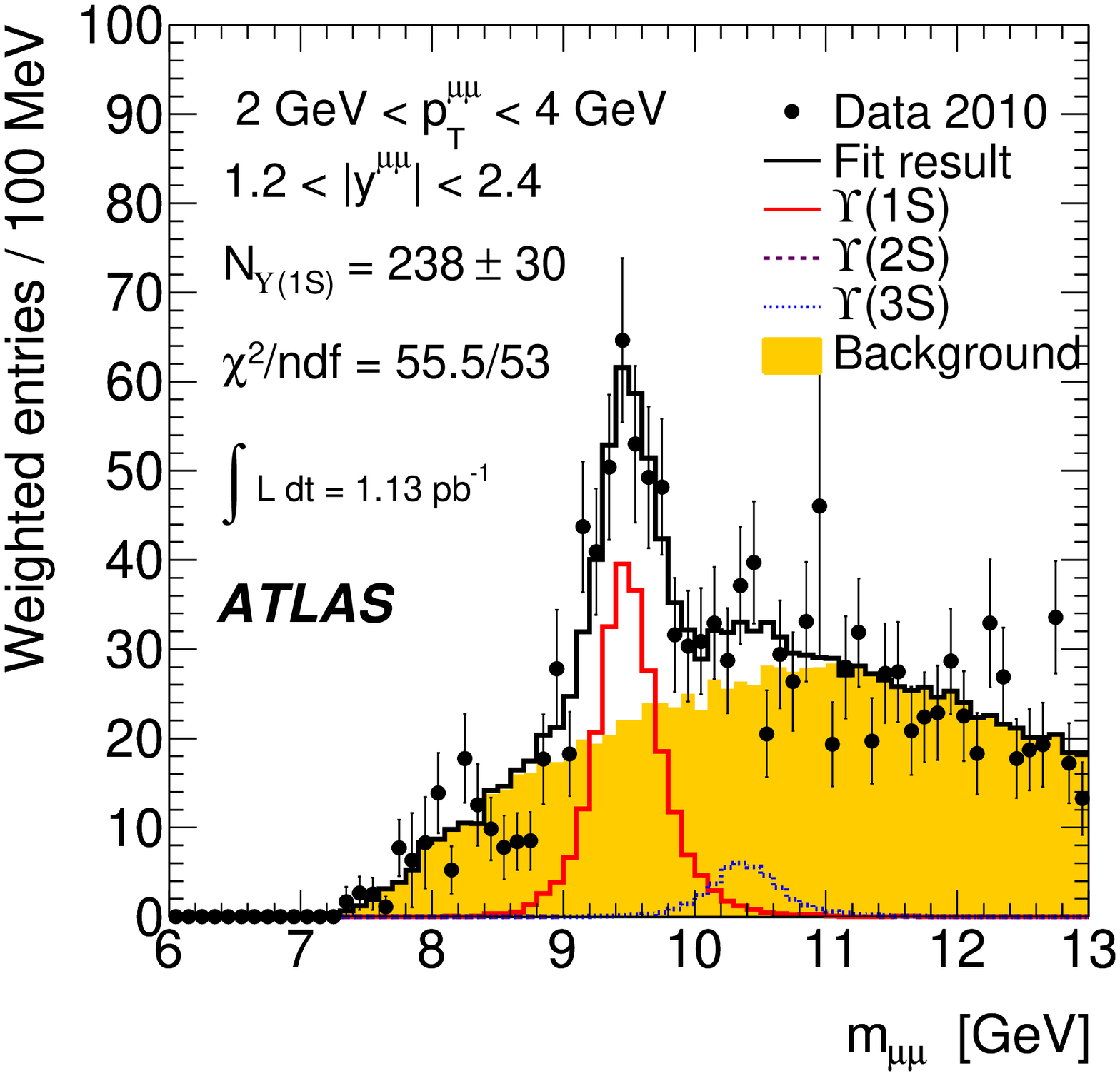}}
\subfigure[]{\includegraphics[width=0.45\textwidth]{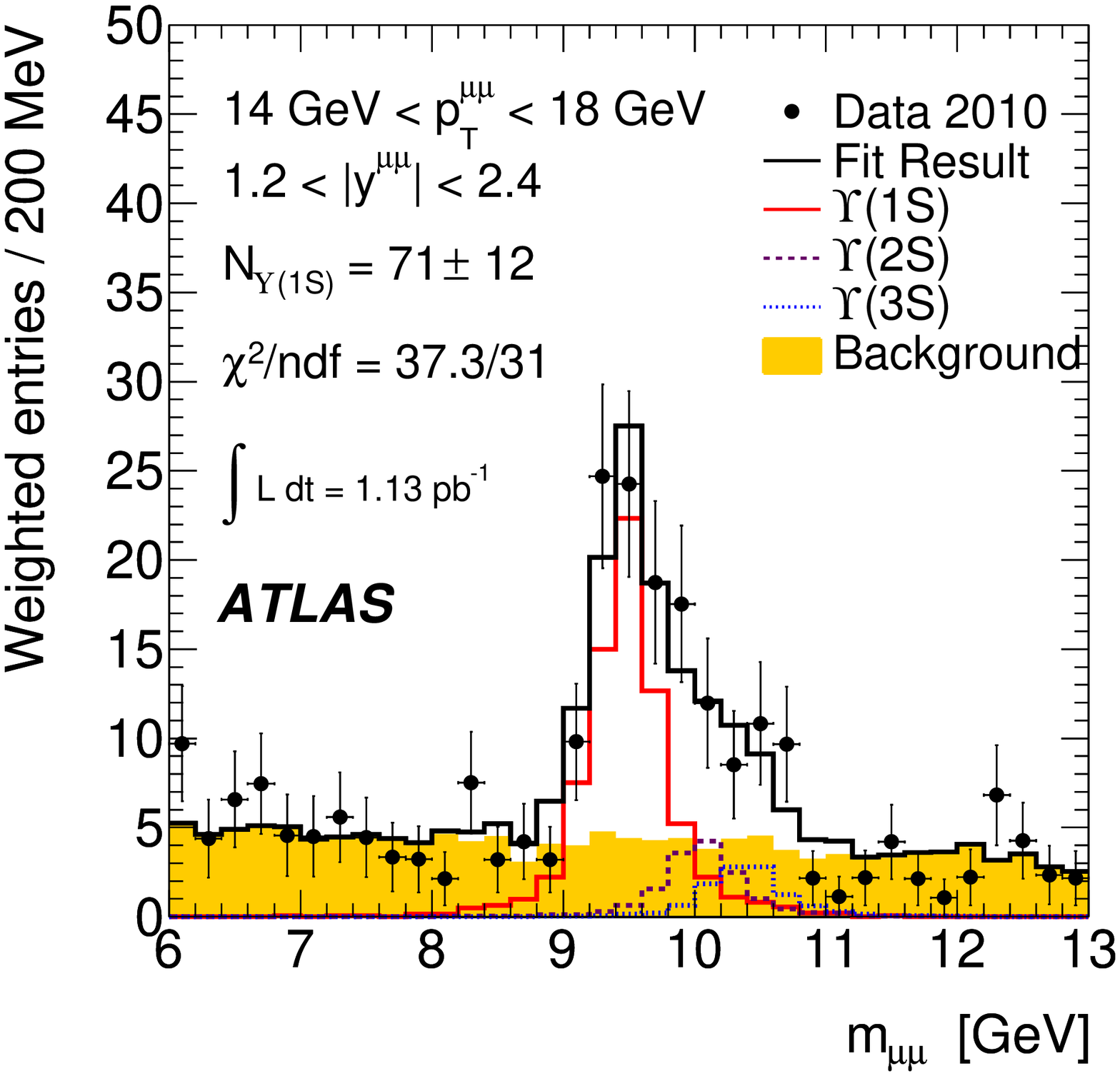}}
\caption{Dimuon mass distributions for four representative bins in $\ymm$ and $\ptmm$. 
The data (filled circles) are shown together with the result of the unbinned maximum likelihood fit (histogram) as explained in the text. The shaded histogram shows the background contribution, and the three other histograms show the contributions from the three $\Upsilon$ states. All histograms are normalised by the factor determined in the fit. In the individual plots, the fitted $N_{\uones}$ yield with its statistical uncertainty, the $\chi^2$ and the number of degrees of freedom are also given. It should be noted that this is simply a binned graphical representation of the fit; the actual fit is unbinned and interpolates the template histograms to obtain the input probability density function. }
\label{fig:upsfit}
\end{center}
\end{figure*}

The number of $\uones$ events is determined from an unbinned maximum likelihood fit to the dimuon
mass distributions in each bin in $\ptmm$ and $\ymm$ where $\ptmm$ ($\ymm$) is the $p_T$ ($y$) of 
the dimuon system. The distributions and corresponding fit results are shown for four representative 
kinematic bins in Fig.~\ref{fig:upsfit}.

The shape of these distributions is rather complex. 
The background varies substantially over the considered mass range
and its shape changes significantly depending on the kinematic bin. At low $\ptmm$,
the background increases sharply with $\mmm$ and is significant in the $\uones$ mass range. 
At high $\ptmm$, the background 
is nearly independent of the dimuon mass and also relatively low compared to the signal. Additionally, the
$\uones$ signal is not well separated from the $\utwos $ and
$\uthrees$, particularly in the forward region ($1.2<|\ymm|<2.4$), due to the limited track 
momentum resolution~\cite{muonmomscale}. Resolving the 
$\utwos $ and $\uthrees$ is even more difficult, and thus measurements of $\utwos$ and
$\uthrees$ production are not presented in this letter.

\begin{sloppypar}
A template based likelihood fit method is employed where four
parameters are fitted independently in each kinematic bin: the numbers of $\uones$, $\utwos$
and $\uthrees$ mesons, and a background normalisation parameter.
The three $\Upsilon$ signal templates are taken from the corresponding $\Upsilon$ MC samples. The
background templates are constructed from data by pairing a muon with a track reconstructed in the ID of 
opposite electric charge (OS) that passes the ID track selection requirements described in Section~\ref{sec:evsel} and the
kinematic requirements of $p_T>4$~GeV and $|\eta|<2.5$. This template (denoted as ``OS $\mu$+track'') gives an adequate description of the background since its shape is primarily determined by the kinematic selection requirements. The $\Upsilon$ signal contamination in this sample is expected to be negligible. 
\end{sloppypar}

The fit results for $N_{\uones}$ are given in Fig.~\ref{fig:upsfit} for four kinematic bins. The goodness of fit is assessed by a $\chi^2$ test comparing the data to the template distributions using the normalisations determined by the likelihood fit to the data. In all bins the $\chi^2$ probability is $>5\%$.

\begin{figure*}[htbp]
\begin{center}
\includegraphics[width=0.45\textwidth]{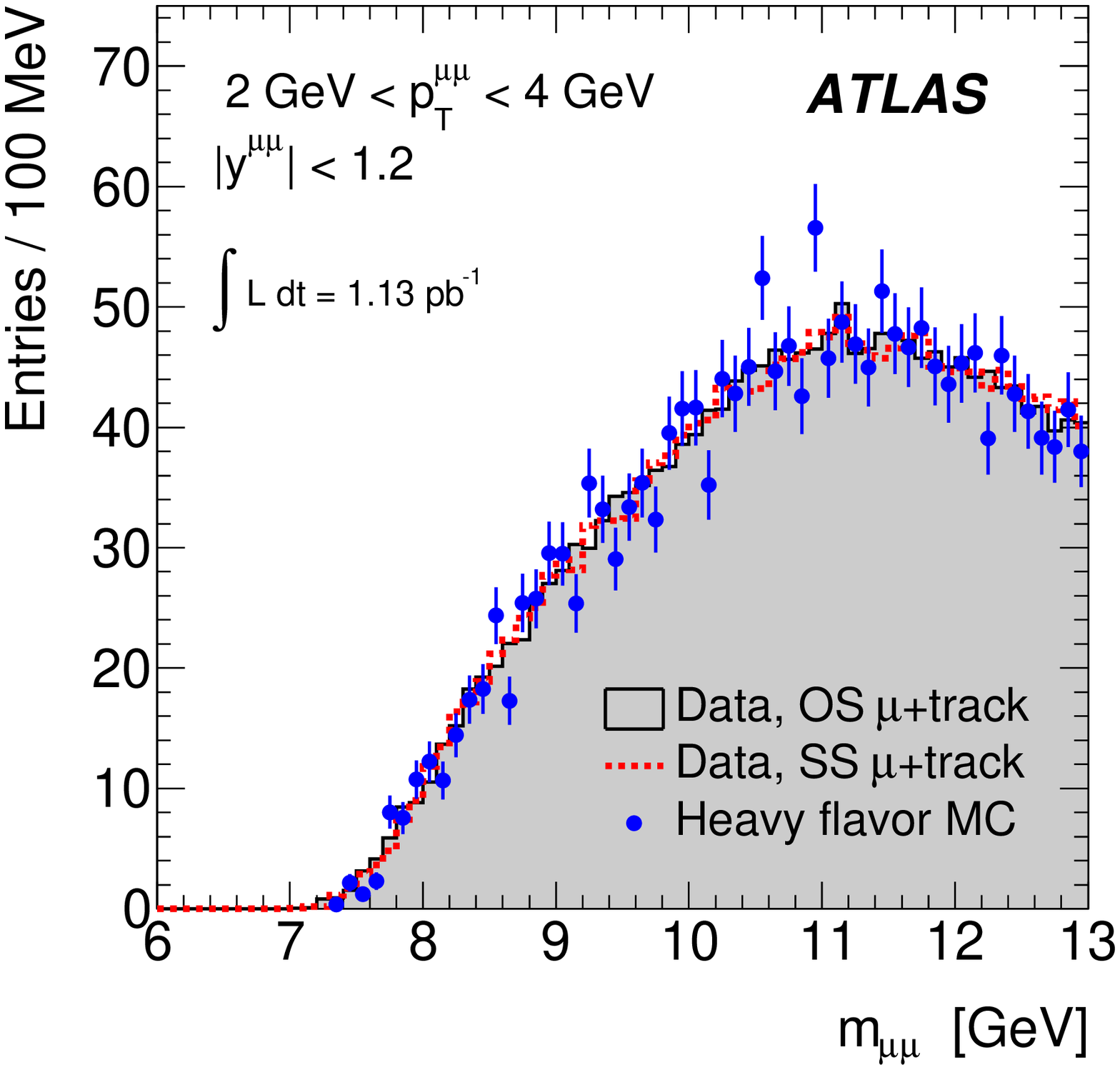}
\includegraphics[width=0.45\textwidth]{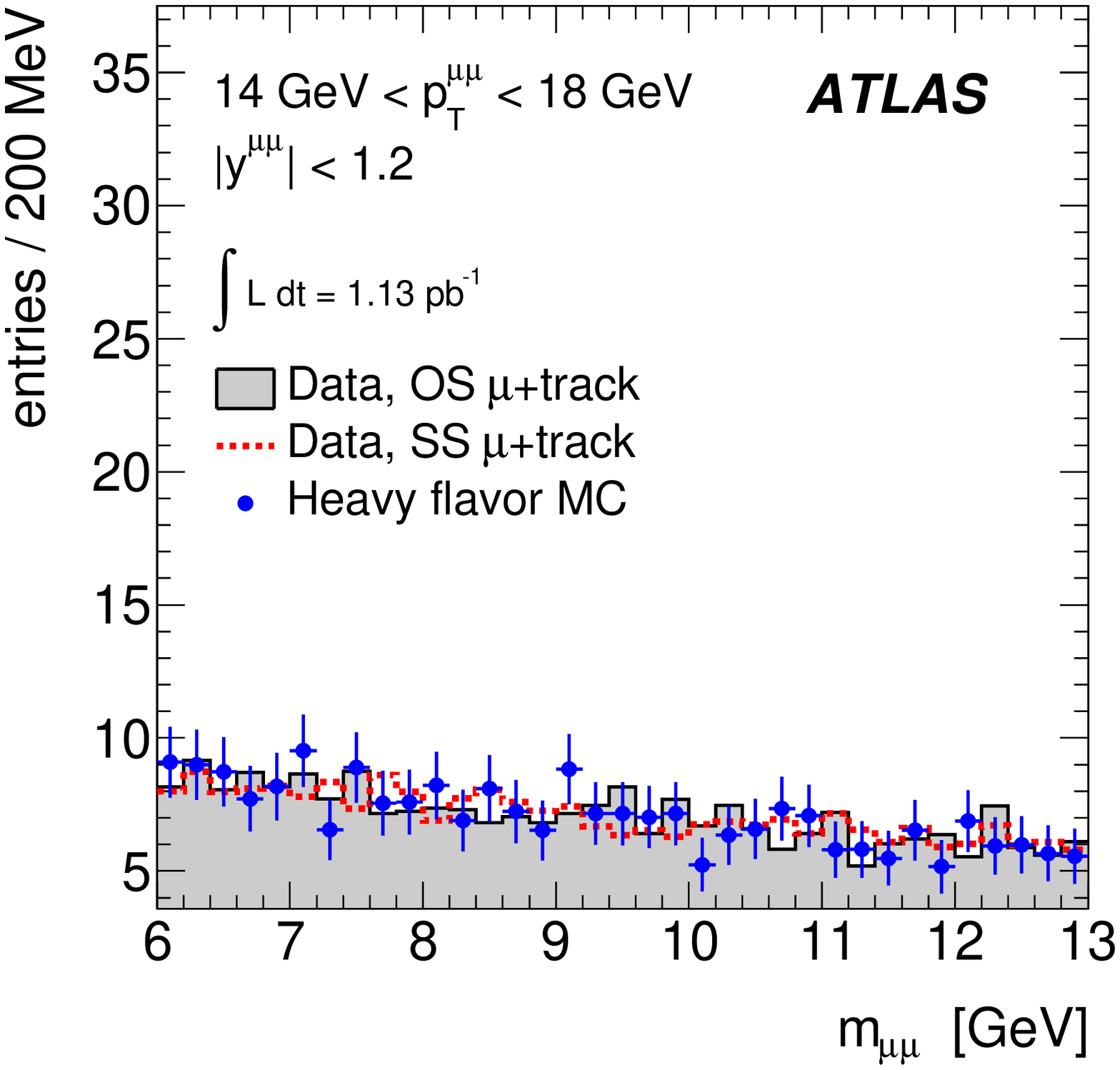}
\caption{Two examples of the templates used for the description of the dimuon mass dependence of the background are shown. 
The solid histogram shows the opposite-sign (OS) $\mu$+track, the dashed histogram shows the same-sign (SS) $\mu$+track and the filled circles show
the histogram derived from dimuon events in open $b\bar{b}$ and $c\bar{c}$ MC (heavy flavour MC). The error bars reflect the statistical uncertainty on the MC-based template. All histograms are normalised to the same absolute amount of background as determined in the fit for each kinematic bin (see also Fig.~\ref{fig:upsfit} and Section~\ref{sec:upsfit}).}
\label{fig:bg}
\end{center}
\end{figure*}

Alternative templates for the shape of the background are constructed using dimuon events in $b\bar{b}$ and $c\bar{c}$ MC or $\mu$+track events with the same electric charge (SS $\mu$+track). A comparison of these alternative background templates shapes is shown in Fig.~\ref{fig:bg} for two representative bins. The shapes of the alternative templates are similar to the default template at both low and high $p_T$, and any differences are considered as part of the systematic uncertainty, as discussed in Section~\ref{sec:sys}. 

The momentum resolution is determined from cosmic rays, $J/\psi$ and $Z$ mass distributions~\cite{muonmomscale} 
and the central mass value for the $\uones$ is fixed to the expected value~\cite{pdgmass}. The validity of fixing the mass resolution and the overall mass scale is considered as part of the systematic uncertainty, as described in Section~\ref{sec:sys}.

\section{Systematic Uncertainties}
\label{sec:sys}

The following systematic uncertainties are considered:
\begin{sloppypar}
The luminosity calibration has been determined using the
van der Meer scan  technique~\cite{vdm} with a precision of 3.4\%~\cite{luminote,lumipaper}. It directly translates into a cross-section uncertainty of 3.4\%.
\end{sloppypar}

\begin{sloppypar}
The uncertainties on the single muon reconstruction efficiency, trigger efficiency~\cite{muoneff} and impact parameter resolution (described in Section~\ref{sec:evsel}) result in 1\%, $<1\%$ and 1-3.5\%  uncertainties on the cross-section measurement, respectively. 

The impact of bin migrations is studied using MC, considering different shapes of the $p_T^{\uones}$ distribution. A systematic uncertainty of 2\% is associated with this.
\end{sloppypar}

\begin{sloppypar}
The uncertainties on the fit model are evaluated using {\it pseudo-experiments}: 1000 sets of pseudo-data are created assuming a Poisson distribution corresponding to the prediction of the default template used for the data fits, and these are then fit with an alternative model as described below. The difference between the mean value of the 1000 pseudo-experiments and the value from the default template is used to assess the systematic effect. This method is used in order to avoid the propagation of statistical uncertainties in the data as part of the systematic uncertainty. However, fitting the data directly with any of the alternative choices gives consistent results. The systematic uncertainties related to the fit arise from the following contributions.
\end{sloppypar}

\begin{itemize}
\item[ -] The uncertainty on the template for the background model is
  assessed  by constructing alternative templates based on the SS $\mu$+track templates and
  on dimuon events in open $b\bar{b}$ and $c\bar{c}$ MC, as described in Section~\ref{sec:upsfit}. 
  The larger of the two differences, as determined using the pseudo-experiment technique described above, is taken as the uncertainty. The resulting uncertainty on the cross-section is typically about 4\% but in some bins can be as high as 8\%. Same-sign dimuon templates were also considered but are statistically too limited.
\item[ -] The uncertainty on the signal model is assessed by changing from the signal templates to Crystal-Ball functions~\cite{crystalball}.
  Here the three $\Upsilon$ states are considered and the differences between their mass values is fixed to the PDG value~\cite{pdgmass}. Their
  relative normalisations are allowed to float as in the default fit. In addition 
  a common scale parameter and the parameters that describe the
  low-mass tail of the Crystal-Ball function float freely in the fit. The width
  is constrained to the value predicted by smeared MC
  to match resolutions measured in data.
  In the central region ($|y^{\uones}|<1.2$) the resulting difference is typically 1\% while in the forward region ($1.2<|y^{\uones}|<2.4$) 
  it is about 5-10\%, and these values are included in the systematic uncertainty. 
  The data were also directly fitted with this alternative model
  and the result was found to be consistent with respect to the default fit.
\item[ -] \begin{sloppypar}
Both the overall mass scale and the separation between the three $\Upsilon$ states are fixed in the fit.
The position of the reconstructed $\uones$ mass peak is influenced by the overall momentum scale and the energy loss. The momentum scale has been determined using
   $J/\psi$ mesons and $Z$ bosons to better than
   0.2\%~\cite{muonmomscale}. The energy loss has been evaluated from the variations of the 
$J/\psi$ and $K_s^0$ masses~\cite{mbpaper} in different regions of the detector.
Evaluated at the world average $\uones$ mass, both effects together allow for a
mass position uncertainty of $25$ MeV.
To assess the corresponding systematic uncertainty, pseudo-experiments
 are used to evaluate the impact of a possible shift in the mass scale by 25~MeV on $N_{\uones}$. 
The difference to the default fit is found to be 
less than 1\% in most bins but up to 5\% in the highest-$p_T$ forward bin. 
The mass separation is only influenced by the momentum scale: considering an uncertainty of 0.2\% 
on the mass scale results in an uncertainty on the cross-section of at most $2\%$ in any bin. 
\end{sloppypar}

\item[ -] The mass resolution is fixed to the value determined from detailed studies based on $J/\psi$s, $Z$s and cosmic rays~\cite{resmuonconfnote}. 
A systematic uncertainty is assessed by changing the momentum resolution by $1\sigma$. The resulting systematic uncertainty on the cross-section measurement is below $3\%$ in all bins.
\end{itemize}

All these systematic uncertainties are uncorrelated and added in quadrature to give the total systematic uncertainty. It is typically 5\% in the central and 10\% in the forward region. This can be compared to the statistical uncertainty of about 10\% (15\%) at low $p_T^{\uones}$ for the central (forward) region and 30\% at the highest $p_T^{\uones}$.

The number of $\uones$ events and the statistical and systematic uncertainties for each bin are listed in Table~\ref{tab:results}.

\begin{table*}[htbp]
\center
\begin{tabular}{c|r|r|rrr}
\hline\hline
$p_T^{\uones}$ & $N_{\uones}$ & ${\rm d}^{2}\sigma/{\rm d}p_{T}{\rm d}y$ &
$\delta_{stat}$& $\delta_{syst}$ & $\delta_{tot}$\\
(GeV) &  & (pb/GeV)  & (\%) & (\%) & (\%) \\
\hline
\multicolumn{6}{c}{}\\
\multicolumn{6}{c}{$|y^{\uones}|<1.2$}\\
\hline
$0-2   $ & 213 & $39.3$ & 12 & 6 & 13 \\
$2-4   $ & 246 & $45.4$ & 10 & 5 & 11 \\
$4-6   $ & 190 & $35.0$ & 12 & 5 & 13 \\
$6-8   $ & 175 & $32.2$ & 12 & 6 & 13 \\
$8-11  $ & 203 & $24.9$ & 10 & 4 & 11 \\
$11-14 $ & 110 & $13.6$ & 13 & 7 & 15 \\
$14-18 $ & 70 & $6.5$ & 15 & 7 & 17 \\
$18-26 $ & 17 & $0.8$ & 33 & 9 & 34 \\
\hline
\multicolumn{6}{c}{}\\
\multicolumn{6}{c}{$1.2<|y^{\uones}|<2.4$}\\
\hline
$0-2   $ & 163 & $30.0$ & 17 & 13 & 22 \\
$2-4   $ & 238 & $43.8$ & 13 & 13 & 18 \\
$4-6   $ & 124 & $22.9$ & 18 & 9 & 20 \\
$6-8   $ & 129 & $23.7$  & 15 & 9 & 17 \\
$8-11  $ & 132 & $16.2$ & 14 & 7 & 16 \\
$11-14 $ & 75 & $9.3$ & 18 & 9 & 20 \\
$14-18 $ & 71 & $6.5$ & 17 & 10 & 20 \\
$18-26 $ & 28 & $1.3$ & 29 & 15 & 33 \\
\hline\hline
\end{tabular}
\caption{Measured differential cross-sections for $\Upsilon_{1S}$ production in the
fiducial acceptance ($p_T^\mu>4$~GeV, $|\eta^\mu|<2.5$ on both muons). Also listed are the corrected yields of $\uones$ mesons,
$N_{\uones}$, and the fractional statistical ($\delta_{stat}$), systematic ($\delta_{syst}$) and total
($\delta_{tot}$) uncertainties. }
\label{tab:results}
\end{table*}

\section{Differential {\boldmath $\uones$} Cross-Section Measurement and Comparison to Theoretical Predictions}
\label{sec:results}

\begin{figure*}[htbp]
\begin{center}
\includegraphics[width=0.49\textwidth]{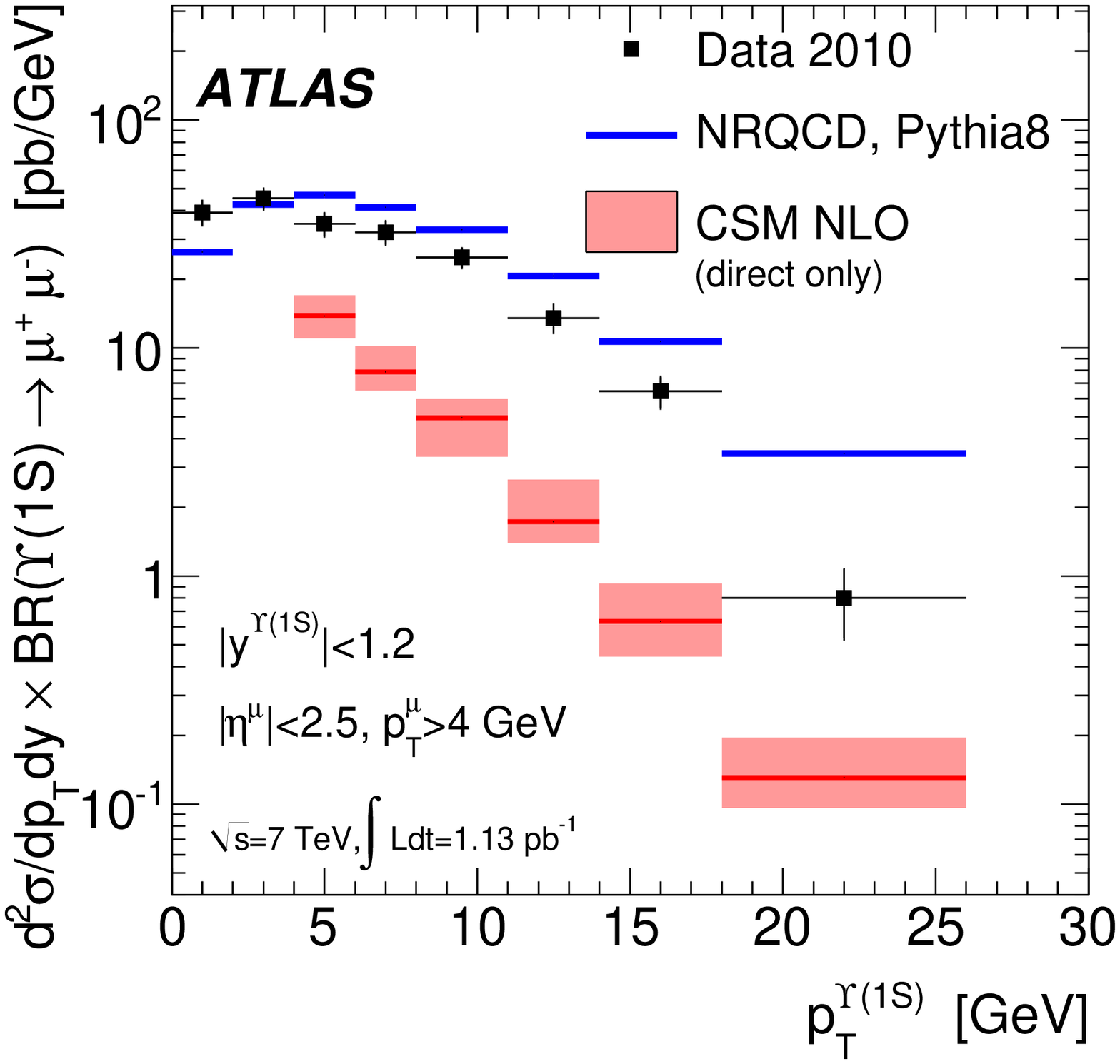}
\includegraphics[width=0.49\textwidth]{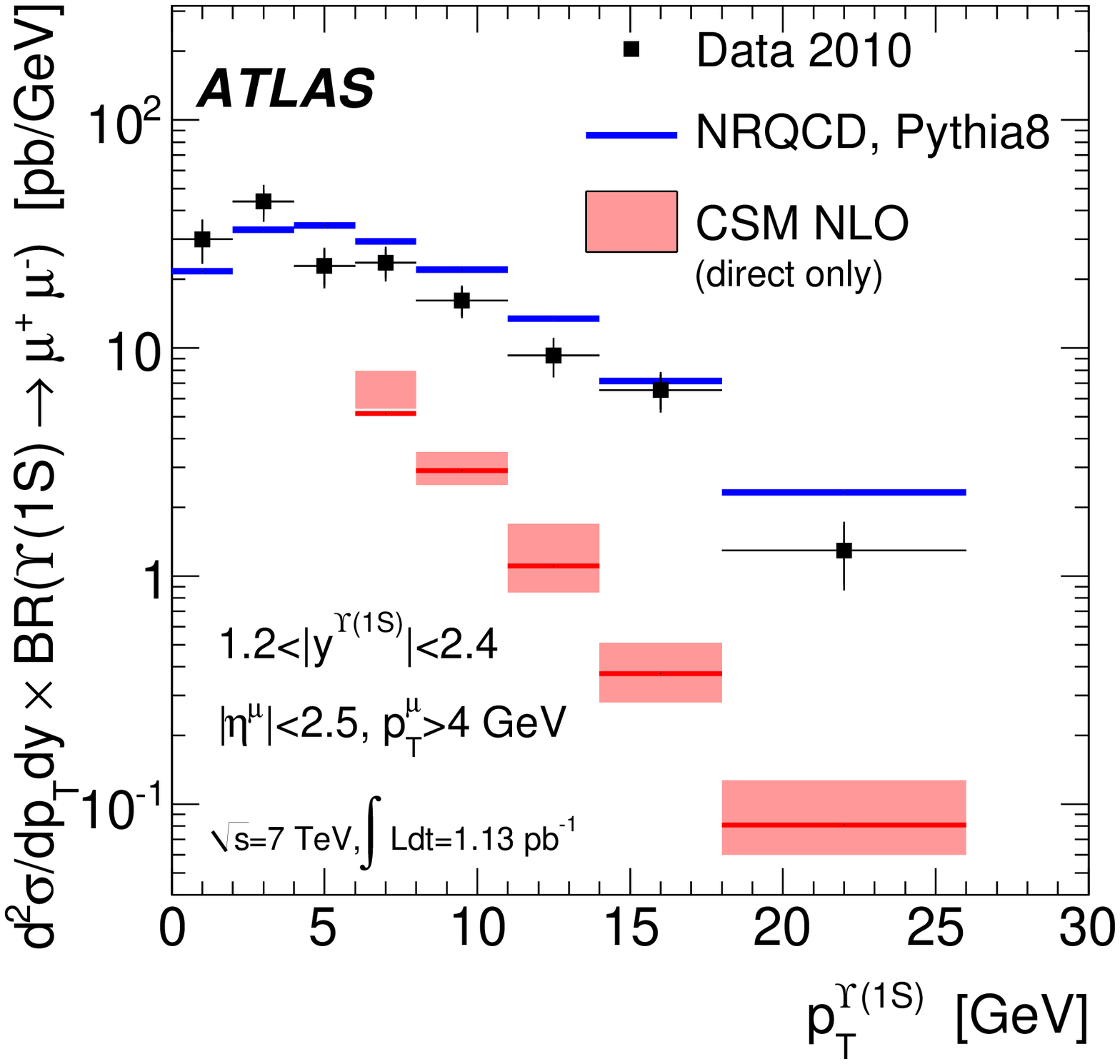}
\caption{The differential $\uones$ cross-section for the $|y^{\uones}|<1.2$ (left) and $1.2<|y^{\uones}|<2.4$ (right) as function of $p_T^{\uones}$
for $p_T^\mu>4$~GeV and $|\eta^\mu|<2.5$ on both muons. 
Also shown is the colour-singlet NLO (CSM) prediction using $m_T$ for the renormalisation and factorisation scales. The shaded area shows the change in the theoretical prediction when varying the renormalisation and factorisation scales by a factor of two.
The CSM NLO calculation accounts only for direct production of $\uones$ mesons and not for any feed-down from excited states.
The NRQCD prediction as implemented in {\sc Pythia8} is also shown for a particular choice of parameters as discussed in the text.}
\label{fig:xsec}
\end{center}
\end{figure*}

\begin{sloppypar}
Figure~\ref{fig:xsec} shows the differential $\uones$ cross-section measurement for the
central and forward regions in 8 bins in $p_T^{\uones}$
covering $0$ to $26$~GeV for $p_T^\mu>4$~GeV and $|\eta^\mu|<2.5$ on both muons. The results are also listed in Table~\ref{tab:results}. 
The cross-section falls by about a factor of 50 over this $p_T$-range in both the central and the forward region.

Two theoretical predictions for quarkonium production are compared to the experimental data in Fig.~\ref{fig:xsec}. 
The first is the result of predictions from {\sc Pythia 8.135}~\cite{py8} using the NRQCD~\cite{nrqcd} framework\footnote{
The following five parameters have been changed from the default {\sc Pythia 8.135} implementation as recommended by the {\sc Pythia} authors: 
{\tt PhaseSpace:pTHatMin=1 GeV}, {\tt PhaseSpace:pTHatMinDiverge=0.5},  {\tt SuppressSmallPT( 1., 3, false)}; 
{\tt Bottomonium:OUpsilon3P08 = 0.02}, {\tt Bottomonium:Ochib03P01=0.085}.}. This prediction
uses the NRQCD matrix elements as recommended in Ref.~\cite{nason} and determined from Tevatron data, and a $p_T$ and $Q^2$ dependent reweighting of the differential cross-section as described in Ref.~\cite{kra08} to prevent a divergence of the cross-section at low $p_T$. The CTEQ5L parton distribution functions are used~\cite{cteq5l}. The second prediction is a next-to-leading-order QCD calculation of the $\uones$ production 
cross-section in a colour-singlet state~\cite{campbell} as implemented in  {\sc MCFM}~\cite{mcfm} 
which corresponds to the leading contribution in the NRQCD expansion (CSM NLO). 
This prediction is not available for $p_T^{\uones}<4$~GeV ($p_T^{\uones}<6$~GeV) in the central (forward) region since 
the perturbative expansion is not under proper control and this fixed-order calculation fails to provide a reliable estimate~\cite{campbell}. Note that this prediction does not reproduce the differential cross-section measurements at the Tevatron unless additional higher-order terms of ${\cal O}(\alpha_s^5)$ are considered and the data are corrected for feed-down from $\chi_b \to \Upsilon\gamma$ production~\cite{tevtheory} and other excited $\Upsilon$ states. The renormalisation and factorisation scales for this prediction are set to $m_T=\sqrt{4m_b^2 + p_T^2}$ where $m_b=4.75$~GeV and $p_T$ is the transverse momentum of the $\uones$. These scales are varied by a factor of two to assess the sensitivity of the calculation to these parameters. The CTEQ6M parton distribution functions are used for this prediction~\cite{cteq6m}.

The measured differential cross-section has a different $p_T$ dependence, but its normalization is in reasonable
overall agreement with the NRQCD prediction. In particular, at high (low) $p_T^{\uones}$ the prediction is systematically higher (lower) than the data. 
The cross-section from data is higher than the NLO colour-singlet prediction over the entire $p_T$ range. Part of the discrepancy can likely be accounted for by the fact that this prediction does not include any feed-down from higher mass states which was estimated to contribute about a factor of 2 at the Tevatron~\cite{feeddown}. In Ref.~\cite{tevtheory}, it was shown that additional higher order corrections increase with increasing $p_T$ (they contribute about a factor of 10 at high $p_T$), and are required to describe the Tevatron data. This is qualitatively consistent with the discrepancies observed between the data presented here and the NLO prediction. Whether or not these data can be consistently understood together with the data from previous collider experiments within a consistent theoretical framework remains to be seen.
\end{sloppypar}

The data are not compared directly to previous measurements~\cite{cdf,d0,cms} since in the present analysis no correction is made for the $p_T^\mu$ and $\eta^\mu$ cuts in order to avoid the need for any assumption on the spin alignment of the $\uones$. 

\section{Conclusions}
\label{sec:conclusions}
A measurement of the differential production cross-section of $\uones$ mesons decaying into two muons measured in the fiducial region with $p_T^\mu>4$~GeV and $|\eta^\mu|<2.5$ has been presented as function of $p_T^\uones$ and $y^\uones$. The typical uncertainty of the measurement is about 10-15\% at low $p_T^{\uones}$ and 35\% at high $p_T^{\uones}$ and is dominated by the statistical precision of the data. Due to the restriction to $p_T^\mu>4$~GeV and $|\eta^\mu|<2.5$, the measurement is almost independent of any assumption on the spin alignment of the $\uones$.

\begin{sloppypar}
The data significantly exceed the NLO prediction but this may well be explained due to contributions from higher mass bound states and by the need for additional higher order corrections to $\uones$ production. In contrast, the data are in reasonable agreement with the NRQCD prediction as implemented in {\sc Pythia8} but differences in the shape of the $p_T$ spectrum of about a factor of two are observed.  The data presented here will be useful to further understand the complex mechanisms that govern quarkonium production.
\end{sloppypar}
 
\begin{sloppypar}
\section{Acknowledgements}

We thank CERN for the very successful operation of the LHC, as well as the
support staff from our institutions without whom ATLAS could not be
operated efficiently. We also wish to thank Pierre Artoisenet, John Campbell and 
Torbj\"orn Sj\"ostrand for their help with the theoretical predictions.

We acknowledge the support of ANPCyT, Argentina; YerPhI, Armenia; ARC,
Australia; BMWF, Austria; ANAS, Azerbaijan; SSTC, Belarus; CNPq and FAPESP,
Brazil; NSERC, NRC and CFI, Canada; CERN; CONICYT, Chile; CAS, MOST and
NSFC, China; COLCIENCIAS, Colombia; MSMT CR, MPO CR and VSC CR, Czech
Republic; DNRF, DNSRC and Lundbeck Foundation, Denmark; ARTEMIS, European
Union; IN2P3-CNRS, CEA-DSM/IRFU, France; GNAS, Georgia; BMBF, DFG, HGF, MPG
and AvH Foundation, Germany; GSRT, Greece; ISF, MINERVA, GIF, DIP and
Benoziyo Center, Israel; INFN, Italy; MEXT and JSPS, Japan; CNRST, Morocco;
FOM and NWO, Netherlands; RCN, Norway; MNiSW, Poland; GRICES and FCT,
Portugal; MERYS (MECTS), Romania; MES of Russia and ROSATOM, Russian
Federation; JINR; MSTD, Serbia; MSSR, Slovakia; ARRS and MVZT, Slovenia;
DST/NRF, South Africa; MICINN, Spain; SRC and Wallenberg Foundation,
Sweden; SER, SNSF and Cantons of Bern and Geneva, Switzerland; NSC, Taiwan;
TAEK, Turkey; STFC, the Royal Society and Leverhulme Trust, United Kingdom;
DOE and NSF, United States of America.

The crucial computing support from all WLCG partners is acknowledged
gratefully, in particular from CERN and the ATLAS Tier-1 facilities at
TRIUMF (Canada), NDGF (Denmark, Norway, Sweden), CC-IN2P3 (France),
KIT/GridKA (Germany), INFN-CNAF (Italy), NL-T1 (Netherlands), PIC (Spain),
ASGC (Taiwan), RAL (UK) and BNL (USA) and in the Tier-2 facilities
worldwide.
\end{sloppypar}

\onecolumn

\clearpage
\begin{flushleft}
{\Large The ATLAS Collaboration}

\bigskip

G.~Aad$^{\rm 48}$,
B.~Abbott$^{\rm 111}$,
J.~Abdallah$^{\rm 11}$,
A.A.~Abdelalim$^{\rm 49}$,
A.~Abdesselam$^{\rm 118}$,
O.~Abdinov$^{\rm 10}$,
B.~Abi$^{\rm 112}$,
M.~Abolins$^{\rm 88}$,
H.~Abramowicz$^{\rm 153}$,
H.~Abreu$^{\rm 115}$
% ...
\bigskip

% ...
$^{10}$ Institute of Physics, Azerbaijan Academy of Sciences, Baku, Azerbaijan\\
$^{11}$ Institut de F\'isica d'Altes Energies and Universitat Aut\`onoma  de Barcelona and ICREA, Barcelona, Spain\\
% ...
$^{48}$ Fakult\"{a}t f\"{u}r Mathematik und Physik, Albert-Ludwigs-Universit\"{a}t, Freiburg i.Br., Germany\\
$^{49}$ Section de Physique, Universit\'e de Gen\`eve, Geneva, Switzerland\\
% ...
$^{88}$ Department of Physics and Astronomy, Michigan State University, East Lansing MI, United States of America\\
% ...
$^{111}$ Homer L. Dodge Department of Physics and Astronomy, University of Oklahoma, Norman OK, United States of America\\
$^{112}$ Department of Physics, Oklahoma State University, Stillwater OK, United States of America\\
% ...
$^{115}$ LAL, Univ. Paris-Sud and CNRS/IN2P3, Orsay, France\\
% ...
$^{118}$ Department of Physics, Oxford University, Oxford, United Kingdom\\
% ...
$^{153}$ Raymond and Beverly Sackler School of Physics and Astronomy, Tel Aviv University, Tel Aviv, Israel\\
% ...
\end{flushleft}


\end{document}